\documentclass[preprint,nofootinbib,12pt]{revtex4}

\usepackage{graphicx}
\usepackage{psfrag}
\usepackage{axodraw}
\usepackage{pstricks}

\newcommand{\GeV}{{\rm GeV}}

\newcommand{\TeV}{{\rm TeV}}
\newcommand{\eV}{{\rm eV}}

\renewcommand{\Im}{\mathcal{I}m}

\newcommand{\beq}{\begin{equation}}
\newcommand{\eeq}{\end{equation}}
\newcommand{\beqa}{\begin{eqnarray}}
\newcommand{\eeqa}{\end{eqnarray}}

\newcommand{\lsim}{\mathrel{\rlap{\lower4pt\hbox{\hskip1pt$\sim$}}
    \raise1pt\hbox{$<$}}}         
\newcommand{\gsim}{\mathrel{\rlap{\lower4pt\hbox{\hskip1pt$\sim$}}
    \raise1pt\hbox{$>$}}}         

\begin{document}


\vspace*{.0cm}

\title{Leptogenesis with Composite Neutrinos}

\author{Yuval Grossman}\email{yuvalg@lepp.cornell.edu}
\author{Yuhsin Tsai}\email{yt237@lepp.cornell.edu}

\affiliation{\vspace*{4mm}Institute for High Energy
Phenomenology\\ Newman Laboratory of Elementary Particle Physics\\
Cornell University, Ithaca, NY 14853, USA\vspace*{6mm}}


\vspace{1cm}
\begin{abstract}
Models with composite singlet neutrinos can give small Majorana or
Dirac masses to the active neutrinos.  The mechanism is based on the
fact that conserved chiral symmetries give massless neutrinos at the renormalizable level. Thus, they acquire very small
masses due to non-renormalizable terms. We investigate such models in
two aspects. First, we find UV completions for them and then we
investigate the possibility of giving leptogenesis. We find that these
models offer new possibilities for leptogenesis. Models with Majorana
masses can exhibit standard leptogenesis. Models with Dirac masses
can provide a realization of Dirac type leptogenesis with mass scale
that can be as low as $10\,\TeV$.

\end{abstract}

\maketitle


\section{INTRODUCTION}

In recent years it has become clear that neutrinos have very small masses
and that they mix.  The origin of these masses is still an open
question. The see-saw mechanism is probably the most elegant
explanation for small neutrino masses. The idea is to add heavy
Majorana right handed (RH) neutrinos to the theory. These added particles give very small Majorana masses to the active, Standard Model (SM) neutrinos.
The see-saw mechanism has one more virtue: it provides an elegant
mechanism to explain the observed baryon asymmetry in the
universe. The idea of this mechanism, called Leptogenesis
(LG)~\cite{Fukugita:1986hr}, is that the heavy RH neutrinos that drive
the see-saw also generate lepton asymmetry when they decay. Part of
this lepton asymmetry is transformed into the observed baryon
asymmetry of the universe (for a review see~\cite{Davidson:2008bu}).

While the see-saw mechanism is very simple and successful, it is not
the only way to explain the observed small neutrino
masses. Another idea for getting light neutrinos that has not been
widely discussed is that of composite RH
neutrinos~\cite{ArkaniHamed:1998pf,Okui:2004xn}. The basic idea is
that there exists a new sector with strong dynamics at a scale
$\Lambda$.  The confinement in this sector leaves some chiral
symmetries exact and produces massless composite fermions.  The only
interaction between the preons of the new sector and the SM sector is
via heavy messengers with large masses of order $M$.  Then, the
Yukawa coupling between the LH and RH neutrinos is suppressed by
powers of the small factor $\Lambda/M$. This can give a natural
explanation for small Dirac or Majorana neutrino masses.

In this article we further investigate the composite RH neutrino
idea. First, we find UV completions for models that give Dirac or
Majorana neutrino masses. We then study how these full models can give
LG. We find that it exhibits interesting LG possibilities. In
particular, it can have see-saw like LG and a low mass scale Dirac LG.

In the next section, we give a brief review of the composite RH
neutrino idea of ref.~\cite{ArkaniHamed:1998pf}. We find UV complete
theories in section III for both Dirac and Majorana neutrinos where
the new particle content is given, and the experimental constraints
are discussed. In section IV, we study LG possibilities in the
model. When the temperature $T$ is below the confinement scale,
$T\ll\Lambda$, and the RH neutrinos are heavy, the composite structure
of the RH neutrinos cannot be probed and standard LG become possible
(IV.A). When $T\sim M\gg\Lambda$, the preons are asymptotically free
and standard LG cannot work. In the case of Dirac neutrinos, the decay
of heavy messengers gives a realization of a low
energy Dirac LG (IV.B). In section V we conclude. A detailed calculation of
the effective couplings is given in Appendix A.  The experimental
bounds on the masses and couplings of the new fields arising from lepton
flavor violating processes are given in Appendices B and C.


\section{COMPOSITE RIGHT-HANDED NEUTRINO}

We first review the idea of composite right-handed
neutrinos~\cite{ArkaniHamed:1998pf}. Consider a new strong sector such
that all the new fields are SM singlets. Like QCD, where the
interaction becomes strong at a scale $\Lambda_{QCD}$, the new sector
becomes strong at a new scale $\Lambda$. Unlike QCD, however, we
assume that the confinement in the new sector keeps some of the chiral
symmetries unbroken. In that case, massless composite fermions are generated
since they are required for anomaly matching of the unbroken chiral
symmetries.

The view point in \cite{ArkaniHamed:1998pf} is that of an effective
field theory where the model is a low energy description of a more
fundamental theory. In that case one needs to include 
non-normalizable operators that are suppressed by some high energy
scale $M$. We can think about such operators as emerging from
integrating out heavy fields. That is, it is assumed that the
``preons'' in the new sector interact with the SM fields through
``messengers.'' The messengers are fields that are charged under both
the SM and the preon sector, and are assumed to be very heavy, with the 
mass scale $M\gg\Lambda$. After confining dynamics occur, the
couplings between the composite fermions and the SM fields are
naturally suppressed by powers of the small ratio $\Lambda/M$. In
particular, the fact that the coupling between the composite and SM
fermions are suppressed makes the composite fermions candidates to be
light RH neutrinos.

The work of Ref.~\cite{Dimopoulos:1980hn} is a well known example of a
model and strong dynamics with unbroken chiral symmetries. The model
is based on an SU$(n+4)_C$ gauge group with a single antisymmetric
tensor $A$ and $n$ antifundamentals $\psi_{f}$ (with $f=1..n$). Below
the confinement scale the theory is described by $n(n+1)/2$
massless composite ``baryons" $\hat B_{ff'}=\hat
B_{f'f}=\psi_{f}A\psi_{f'}$. These baryons are identified with the RH
neutrinos.

In this work, we focus on the $n=2$ case, that is a model with a gauge
group SU$(6)_C$. This model has three massless baryons that can give
mass to the three SM neutrinos.  These baryons are connected to the SM
neutrinos through higher dimension operators suppressed by the high
mass scale $M$. The lowest dimension operator of interest is
\beq\label{eq1}
\lambda^{ff',i}\frac{(\psi^T_fA^*\psi_{f'})L^{\dagger}_{i}\tilde{H}}{M^3}\equiv\lambda^{ff',i}\epsilon^3
{B}_{ff'}L^{\dagger}_{i}\tilde{H},
\eeq 
where $i=1,2,3$ runs over the three SM generations and we define
\beq
\epsilon\equiv {\Lambda \over M}, \qquad
{B}_{ff'}\equiv\frac{\psi^T_fA^*\psi_{f'}}{\Lambda^3}, \qquad
\tilde{H}\equiv i\sigma^2H^*,
\eeq 
such that ${B}_{ff'}$ are the canonically normalized baryon fields. If
lepton number is a good symmetry of the model, the term in (\ref{eq1})
generates Dirac masses to the SM neutrinos
\beq \label{d-mass}
m_{\nu}=\lambda\epsilon^3v,
\eeq
where $v$ is the Higgs vev and flavor indices are suppressed.

We can also include lepton number violating terms in the theory. Then
we have the well known see-saw term
\beq
y_{ij}{\bar{L}_i\bar{L}_jHH \over M}.
\eeq
In addition, there are new terms involving the composite fermions
\beq\label{eq2} 
h^{ff',gg'}\frac{(\psi_f A\psi_{f'})(\psi_g
A\psi_{g'})}{M^5}=h^{ff',gg'} M\epsilon^6B_{ff'}B_{gg'}.
\eeq 
The neutrino mass matrix is now a $6 \times 6$ matrix that in the
$({L_{\alpha},B_{ff'}})$ basis is given by
\beq
\left[\begin{array}{cc}  y v^2/M& \lambda\epsilon^3v \\ \lambda\epsilon^3v &
h\epsilon^6M \\
\end{array}\right],
\eeq 
where flavor indices are implicit.  Diagonalizing the matrix and
assuming that all the dimensionless couplings are order one we get
\beq\label{eq4}
m_{\nu}\sim\frac{v^2}{M} ,\qquad m_{N}\sim \epsilon^6
M ,\qquad \theta_{LR}\sim
\min\left(\,\sqrt{\frac{m_{\nu}}{m_N}},\sqrt{\frac{m_N}{m_{\nu}}}\,\right).
\eeq 
$m_{\nu}$ and $m_N$ are, respectively, the LH and RH neutrino
masses, and $\theta_{LR}$ are the mixing angles between the LH and RH
neutrinos. 

We learn that composite RH neutrinos can naturally give small neutrino
masses. They can be Dirac masses, eq. (\ref{d-mass}), or Majorana
masses, eq. (\ref{eq4}).

\section{THE UV COMPLETE THEORY}

In~\cite{ArkaniHamed:1998pf} a low energy effective theory approach
was used. In this section, we give UV completions of the models
studied in~\cite{ArkaniHamed:1998pf}.  In III.A, we present the
particle content.  In III.B, the interactions relating to the new
fields are listed and the number of physical parameters is
discussed. In III.C, we obtain bounds on the parameters from $\mu \to
e \gamma$ and muon-conversion experiments. In Appendix.~A, we show how
the coupling of eqs. (\ref{eq1}) and (\ref{eq2}) are obtained by integrating
out the heavy fields of the UV complete theory.

\subsection{Particle Content}

\begin{table}[t]
\begin{center}
\begin{tabular}[t]{c|cccccccc} \hline
& SU(6)$_C$ & SU(2)$_L$ & $U(1)_Y$ & $Q$ & spin & $L$ & $Q_{ps}$ & $SU(2)_{\psi}$
\\ \hline 
$_i L^{\alpha}_L$      & $1$   & $2$   & $-\frac{1}{2}$
& $0$, $-1$ & $\frac{1}{2}$ & $1$ & $0$ & $1$\\ 
$_i E_R$      & $1$   & $1$   & $-1$
& $-1$ & $\frac{1}{2}$ & $-1$ & $0$& $1$\\ 
$H^{\alpha}$ & $1$ & $2$ & $\frac{1}{2}$ & $1$, $0$ & $0$ & $0$ & $0$ & $1$\\ \hline
$_{g}\Omega_{ab}^{\alpha}$ & $15$  &  $2$ & $-\frac{1}{2}$ & $0$,
     $-1$ & $0$ & $0$ & $2$& $1$\\ \hline
$_f\psi_a$     & $6$ & $1$  & $0$ & $0$   &$\frac{1}{2}$&$0$&$1$& $2$ \\
$A_{ab}$  & $15$  &  $1$ & $0$    & $0$  &$\frac{1}{2}$ &$-1$&$2$  & $1$\\
$\Phi_{ab}$   & $15$   &  $1$ & $0$    & $0$  &$0$ &$0$&$2$ & $1$ \\ \hline
$_{k}N$  & $189$ ; $1$ &  $1$ & $0$ & $0$ &$\frac{1}{2}$ & break & $0$ & $1$ \\
   \hline
\end{tabular}
\end{center}
\caption{
The fermions and scalars of the SU(6)$_C$ model. We divide the
particles into four groups. From top to bottom: the SM fields, the
messenger, the preons and the optional lepton number violating
Majorana fermion.\label{table-fields}}
\end{table}

We consider the case of an SU(6)$_C$ gauge symmetry in the preon
sector. As we mention before, this gives three composite neutrinos. The generalization for models with a larger symmetry is
straightforward. The minimum particle content of this model is listed
in Table \ref{table-fields}. In the table we identify representations
by their dimension. In the SM sector, $_iL_L^{\alpha}$ and
$H^{\alpha}$ are lepton and Higgs doublets carrying SU(2)$_{L}$ index
$\alpha=1,2$ while $_iE_R$ is an SU(2)$_L$ singlet. $L$ and $E$ carry
generation index $i=1,2,3$.

There are two types of fermions in the preon sector. The first fermion, $_f\psi_a$, is a fundamental under SU(6)$_C$ that carries a flavor index $f=1,2$ and SU(6)$_C$ index, $a=1,2,...,6$. The other fermion, $A_{ab}$, is a
second rank antisymmetric tensor, that is it belongs to the
$(0,1,0,0,0)$ representation of SU$(6)$. Composite fermions are made
of these two types of fundamental fermions.

Aside from the fermions we also need scalars that connect the fermions to
the SM fields. One scalar, $_{g}\Omega_{ab}^{\alpha}$, is a heavy
messenger, as it is charged under both the SM and preon gauge
groups. It carries a generation index $g=1,2$ (as discuss below, this is
necessary for LG) and transforms as a second rank antisymmetric tensor
under SU(6)$_C$ and as a fundamental under SU(2)$_L$. The other heavy
scalar, $\Phi_{ab}$, used for connecting two $\psi$'s together,
transforms as a second rank antisymmetric tensor under SU(6)$_C$. The
mass scale of both heavy scalars is $M$, which is assumed to be much
larger than the preon confinement scale $\Lambda$.

Lastly, in models with lepton number violation we need one more field
that breaks lepton number. This field, $_{k}N$, is a SM singlet, and
can be either a singlet or a $189$ of SU(6)$_C$.  [The 189 of SU(6) is
$(0,1,0,1,0)$.] Here $k=1,2$ is the generation index, which is needed,
as discuss below, for LG.

There are three accidental symmetries for this model, U(1)$_L$,
U(1)$_{ps}$, and SU(2)$_\psi$. U(1)$_L$ is the SM lepton number
$L$. It is exact in the model without $N$, but broken when 
the Majorana field $N$ is included.  U(1)$_{ps}$, where ``ps" stands for ``preon
sector'', corresponds to a preon sector charge, $Q_{ps}$. Only
preons and heavy scalars carry such charge.  SU(2)$_\psi$ is a
symmetry due to the antisymmetry of the $\psi$ field and correspond to
flavor rotation between the two flavors of $\psi$. Only $\psi$ is
charged under this symmetry.

\subsection{Interactions}

We move to discuss the renormalizable interaction terms of the
model. The SM Yukawa interactions
\beq \label{SM-yuk}
Y^e_{ij}\bar{L}_L^iHE^j_{R}+h.c.,\qquad i,j=1,2,3,
\eeq
are well known, and we do not discuss them any further. We only recall
that the Yukawa couplings, $Y^e_{ij}$, contain 9 complex parameters.

There are mass terms for the new scalar fields
\beq \label{scal-mass}
M_{\Omega gg'}^{2}\Omega^{\dagger}_g\Omega_{g'}+
M^2_{\Phi}\Phi^{\dagger}\Phi.
\eeq
Here $M^2_{\Phi}$ is a dimensionfull coupling with 1 real parameter,
and $M^2_{\Omega}$ is a $2\times 2$ hermitian matrix with 3 real and 1
imaginary parameters. We assume that all new masses are of the same
order, $M^2_\Omega\sim M^2_\Phi\sim M^2$.

There are also interaction terms that involve the new fields. In both
the L-conserving and L-violating models, the following terms are the
most relevant to our study
\beqa \label{eq5} 
Y^L_{gi} A\Omega^{\dagger}_{g}L_{i}&+&h.c.,
\\
\label{eq6}
\tilde{M}_{g}\tilde{H}^{\dagger}\Phi^{\dagger}\Omega_{g}&+&h.c.,
\\
\label{eq7} Y^A_{ff'}\psi_{f}\Phi^{\dagger}\psi_{f'}&+&h.c. \,.
\eeqa
These couplings generate the effective Yukawa interaction of
(\ref{eq1}) via the diagram in Fig.~\ref{fig:1}a (see appendix A). The
coupling $Y^L_{gi}$ is a general $2\times 3$ matrix containing 6 real
and 6 imaginary parameters. $\tilde{M}_{g}$ corresponds to two
dimension full complex coefficients with $g=1,2$. We assume that each of
the elements of $\tilde{M}_{g}$ is of order $M$. The coupling
$Y^A_{ff'}$ is a $2\times 2$ antisymmetric matrix with 1 complex
parameter (see appendix A). 

In the L-violating case we include the $N$ field. The
relevant couplings include a Majorana mass term 
\beq \label{N-mass}
M_{Nkk'} N_kN_{k'},
\eeq
where we assume $M_N \sim M$, and interaction terms
\beq \label{eq8} 
Y^N_{k}\Phi^{\dagger}AN_{k}+h.c.\,.
\eeq 
The mass term (\ref{N-mass}) and the interaction term (\ref{eq8}) are
included for the two possible representations of $N$, the singlet and
the $189$. These two terms generate the L-violating term of eq.~(\ref{eq2})
through the diagram in Fig.~\ref{fig:1}b. If $N$ is a singlet
under all the gauge symmetries, an additional coupling
\beq\label{eq9}
y^N_{ik}H^{\dagger}L_{i}N_{k}+h.c.,
\eeq 
exists. This term is the usual Yukawa coupling in the see-saw
mechanism. Together with the mass term of (\ref{N-mass}) it generates
the usual see-saw term for the light neutrinos.


Aside from the couplings relating to neutrino masses and LG, there are
couplings that connect the new scalars to the SM Higgs field
\beq
\label{eq10}
\lambda^{\Omega(1)}_{gg'}H^{\dagger}\Omega_g H^{\dagger}\Omega_{g'}
+h.c., \qquad
\lambda^{\Omega(2)}_{gg'}H^{\dagger}H\Omega^{\dagger}_g\Omega_{g'}, \qquad
\lambda^{\Phi}H^{\dagger}H\Phi^{\dagger}\Phi.
\eeq
These couplings result in having a Higgs mass much above the weak
scale unless they are fine-tuned. This is the usual fine tuning
problem of the SM. In this work we do not try to solve this problem,
we just assume that there is a solution. Thus, in the following we
assume that the couplings in (\ref{eq10}) vanish.

\begin{figure}[t]
\centering
\begin{picture}(359,89) (15,-65)
    \SetWidth{0.5}
    \ArrowLine(85,18)(138,18)
    \DashArrowLine(50,-27)(85,-27){5}
    \ArrowLine(23,-9)(50,-27)
    \ArrowLine(23,-45)(50,-27)
    \ArrowLine(85,18)(23,18)
    \DashArrowLine(85,-25)(85,18){5}
    \DashArrowLine(138,-27)(85,-27){5}
    \Text(147,18)[lb]{\normalsize{\Black{$L$}}}
    \Text(147,-27)[lb]{\normalsize{\Black{$H$}}}
    \Text(58,-40)[lb]{\normalsize{\Black{$\Phi$}}}
    \Text(13,-9)[lb]{\normalsize{\Black{$\psi$}}}
    \Text(13,-44)[lb]{\normalsize{\Black{$\psi$}}}
    \Text(13,18)[lb]{\normalsize{\Black{$A$}}}
    \Text(95,-9)[lb]{\normalsize{\Black{$\Omega$}}}
    \Text(95,-60)[lb]{\normalsize{\Black{$(a)$}}}
    \ArrowLine(260,2)(277,2)
    \ArrowLine(294,2)(277,2)
    \DashArrowLine(244,-22)(260,2){5}
    \DashArrowLine(311,-22)(294,2){5}
    \ArrowLine(218,-14)(244,-22)
    \ArrowLine(336,-14)(311,-22)
    \ArrowLine(260,2)(218,19)
    \ArrowLine(294,2)(336,19)
    \ArrowLine(218,-40)(244,-22)
    \ArrowLine(336,-40)(311,-22)
    \Text(210,28)[lb]{\normalsize{\Black{$A$}}}
    \Text(344,28)[lb]{\normalsize{\Black{$A$}}}
    \Text(210,-14)[lb]{\normalsize{\Black{$\psi$}}}
    \Text(210,-39)[lb]{\normalsize{\Black{$\psi$}}}
    \Text(344,-14)[lb]{\normalsize{\Black{$\psi$}}}
    \Text(344,-39)[lb]{\normalsize{\Black{$\psi$}}}
    \Text(262,-13)[lb]{\normalsize{\Black{$\Phi$}}}
    \Text(287,-13)[lb]{\normalsize{\Black{$\Phi$}}}
    \Text(273,8)[lb]{\normalsize{\Black{$N$}}}
    \Text(273,-60)[lb]{\normalsize{\Black{$(b)$}}}
\end{picture}
\caption{The diagrams that generate the effective couplings of the model. $(a)$
generates the Yukawa coupling of eq. (\ref{eq1}) and $(b)$ the
L-violating term of eq. (\ref{eq2}).}
\label{fig:1}
\end{figure}
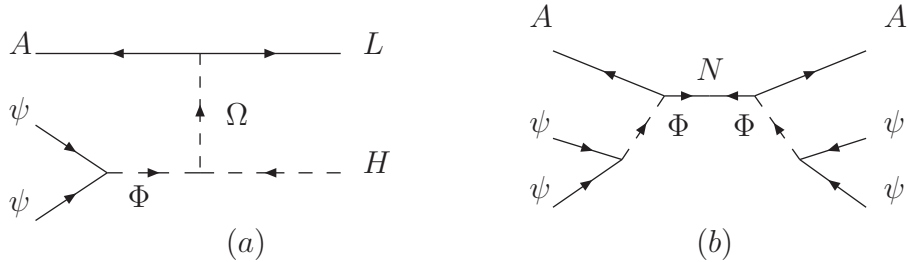

Next we count the number of physical parameters in the various
models. In particular, it is important to show that there are CP
violating phases in the couplings that we used for LG. We start with
the L-conserving model. The parameters of the model discussed above
introduced 22 real and 19 imaginary parameters. The counting is
summarized in Table~\ref{par-count}. Not all of these parameters,
however, are physical. In order to count the number of physics
parameters we need to see how many global symmetries are broken by the
new terms. The global symmetry breaking pattern is
\beqa
U(3)_L\times U(3)_E\times U(1)_{A}\times U(2)_{\psi}\times U(2)_{\Omega}\times
U(1)_{\Phi}\rightarrow U(1)_L\times U(1)_{ps}\times SU(2)_{\psi}.\nonumber
\eeqa 
Thus, we can eliminate 7 real and 16 imaginary parameters
corresponding to the broken generators.  This leave us with 15 real
and 3 imaginary parameters. It is convenient to work in a basis where
all mass parameters are real and diagonal. In that basis the three CPV
phases are in $Y^L$. Note that if we had only one generation for
$\Omega$ there would be no CPV in the model.

\begin{table}[t]
\begin{center}
\begin{tabular}[t]{c|c|c} \hline
 Symbol~ & ~Number of~& ~Number of~ 
\\ & ~parameters (R+I)~& ~Physical parameters (R+I)  
\\ \hline  
$M^2_\Omega$  & 3+1 & 2+0 \\
$M^2_\Phi$ & 1+0 & 1+0 \\
$\tilde M$ & 2+2 & 2+0 \\
$Y^e$ & 9+9 & 3+0 \\
$Y^L$ & 6+6 & 6+3 \\
$Y^A$ & 1+1 & 1+0 \\
\hline
$M_N$  & 3+3 & 2+0 \\
$Y^N$  & 2+2 & 2+1 \\
\hline
$y^N$  & 6+6 & 6+6 \\
\hline
\end{tabular}
\end{center}
\caption{Parameter counting. We divide the couplings into three groups: For the
L-conserving model, we only have the couplings in the first group. For
the L-violating model, if $N$ is a $189$, we have the couplings in
both the first and the second group. When $N$ is a singlet, we have
all the three groups. For each coupling we list the number of
parameter as well as the number of parameter in our ``physical'' basis
choice. We list separately the number of real and imaginary
parameters.
\label{par-count}}
\end{table}

When including the $N$ field there are more parameters and two more
broken global symmetries, U(2)$_N$ and U(1)$_L$. The global symmetry
breaking pattern becomes
\beqa
U(3)_L\times U(3)_E\times U(1)_{A}\times U(2)_{\psi}\times U(2)_{\Omega}\times
U(1)_{\Phi}\times U(2)_N\rightarrow U(1)_{ps}\times SU(2)_{\psi}.\nonumber
\eeqa 
We then eliminate 8 real and 20 imaginary parameters corresponding to
the broken generators. When $N$ is a $189$, there are 19 real and 4
imaginary parameters in the theory. When $N$ is singlet, the model has
25 real and 10 imaginary parameters.

\subsection{Experimental Bounds}

One potential issue with the full model is the contributions of the
heavy particles to rare processes. The effect of new SM singlets is
quite small as they do not couple to SM fields. The messenger,
however, can have significant effect as it charged under the SM gauge
group. Here we study the most significant bounds. They arise from $\mu
\to e\gamma$, muon electron conversion in nuclei, and cosmology.

Starting with $\mu \to e\gamma$, see Fig.~\ref{fig:5}. In the appendix
we calculate the decay rate, eq.~(\ref{muegamma}),
\beqa
Br(\mu\rightarrow\gamma e)=\frac{\alpha |Y^L|^4}{3072\pi G^2_{F}M^4}
\eeqa
Comparing it to the experimental bound \cite{Amsler:2008zz}
\beqa
Br (\mu\rightarrow eX)<1.2 \times 10^{-11},
\eeqa
we obtain a lower bound
\beq \label{eq12} 
M> 10|Y^L|\;\TeV.
\eeq

For coherent muon electron conversion in nuclei (Fig.~\ref{fig:6}), the
theoretical expression is estimated in the appendix, eq. (\ref{eq47}),
\beqa
Br(\mu\rightarrow e,
Ti)\equiv\frac{w_{conv}}{w_{cap}}\approx10^{8}|Y^L|^4\left(\frac{m_{\mu}}{M}\right)^4.
\eeqa
The
experimental bound on the branching ratio is \cite{Kaulard:1998rb}
\beqa
Br(\mu\rightarrow e)<1.7\times 10^{-12}.
\eeqa
Comparing the theoretical prediction with the experimental data
we get a bound on $M$ 
\beq
\label{eq12-again} M > 10|Y^L|\;\TeV.
\eeq
which is the same as the one we get from $\mu \to e \gamma$, (\ref{eq12}).

Aside from the constraints coming from particle physics, constraints
from big-bang nucleosynthesis (BBN) can be strong when the RH
neutrinos have Dirac masses. The reason for this is that the three extra
light modes can be populated before BBN.  Then the energy density,
which depends on the number of relativistic particles, would be
different from the SM one. This difference affects the observed ratio of
primordial elements.

The number of light degrees of freedom is parameterized by the number
of neutrinos. The most stringent bound coming from BBN and CMB data
implies $N_{\nu}\leq 3.3$ at $95\%$ CL
\cite{Cyburt:2004yc}, that is, the effective contribution of the RH
neutrinos can account for as much as $0.3$ of one active neutrino.

This bound rules out any model where the RH neutrinos are populated
at the same temperature as the SM ones. Yet, if the temperature of the
RH sector is lower, the model is viable. The point is that the
contribution to the energy density scales like $T^4$ (where $T$ is the
temperature). Explicitly, the energy density of the SM sector (with
temperature $T_{SM}$) and the three light composite neutrinos (with
temperature $T_{CN}$) is given by \cite{Kolb:1990vq}
\beq
\rho=\frac{\pi^2}{30}(\,g_{*}T^{4}_{SM}+\frac{7}{8}\times 3\times
2\times T^{4}_{CN}\,),
\eeq
where $g_*\simeq 11$ is the effective number of degrees of freedom in
the SM sector (including three massless LH neutrinos).  Requiring that
the RH neutrinos contribute less than $0.3$ active neutrinos is
equivalent to the condition
\beq\label{eqBBN}
3 T^4_{CN} \lsim 0.3 T_{SM}^4\quad \Rightarrow 
T_{CN}\lsim 0.5\, T_{SM}.
\eeq
We learn that we need the composite neutrino temperature to be less
than about half of the SM one in order to satisfy the energy density
constraint from BBN.

Next we compare the temperature of the two sectors.  The preon
confinement scale, $\Lambda$, is larger than the EW scale. Therefore,
the light composite neutrinos decouple from the thermal bath at $T\sim
\Lambda$ which is before the EW phase transitions. Thus, the
temperature of the composite neutrinos is different than that of the
active one. The temperatures ratio is inversely proportional to the
ratio of scale factors, $T_{CN}=(a_i/a_f)\,\Lambda$. The temperature
in the SM sector, however, is not just inversely proportional to the scale
factor, but is higher than this due to the decrease in the number of
degrees of freedom. The total number of degrees of freedom in SM sector is
$g_*\simeq 106$ when $T=\Lambda$ but becomes $g_*\simeq 11$ when
$T=T_{SM}$ just before BBN. Making the conservative assumption that
the EW phase transition is of second order and thus gives no latent
heat, the equality between the initial and the final entropies in SM
sector gives
\beq
106\times a^3_i\times \Lambda^3 = 11\times a^3_f\times T^3_{SM}
\,\Rightarrow\,T_{CN}\simeq 0.47\,T_{SM},
\eeq
which satisfies the BBN bound (\ref{eqBBN}). When the SM is extended
to include extra fields (like in the MSSM) or when the EW phase
transition is first order, $T_{CN}/T_{SM}$ is even smaller and thus
also satisfies the BBN bound.

\section{LEPTOGENESIS}

As has been discussed, one phenomenological use of the composite
model is the realization of leptogenesis. In this section we discuss
two LG possibilities corresponding to different reheating temperatures
and particle contents. First, we study a model with L-violating
interactions and low reheating temperature, $T$, that is,
$T\ll\Lambda$. In this model, standard LG from decays of the heavy
composite RH neutrinos is possible. Second, we study a Lepton number
conserving model with $T\gg\Lambda$. We can have a realization of
Dirac type LG where the new fields can be as light as $10\;\TeV$.

\subsection{Standard leptogenesis}

Consider the L-violating model with $T\ll\Lambda$. In this case, the
preon sector is in its confining phase, and the effects of the
interior structure of the RH neutrinos cannot be probed. The model
looks like the standard see-saw model, and thus we should check if we
can get standard LG in that case.

Using Eq.~(\ref{eq4}), assuming that all dimensionless couplings are
$O(1)$, and setting the active neutrino mass to
$m_{\nu}\sim10^{-2}\;\eV$, the composite RH neutrino mass is of order
\beq\label{eq12.5}
m_N\sim 10^{15}\epsilon^6\;\GeV.
\eeq
We define the standard two parameters~\cite{Davidson:2008bu}
\beqa\label{eq13}
\tilde{m}\equiv8\pi\frac{v^2}{m^2_N}\Gamma_D,\qquad
m_{*}\equiv8\pi\frac{v^2}{m^2_N}H\Big|_{T=m_N}.
\eeqa
They represent the particle decay and the universe expansion rate
relating to LG. The baryon asymmetry is estimated~\cite{Davidson:2008bu}
\beq\label{eq14}
Y_{\Delta B}\simeq
\frac{135\zeta(3)}{4\pi^4g_*}\displaystyle\sum_{\alpha}
\varepsilon_{L\alpha\alpha}\times\eta_{\alpha}\times
C\simeq 10^{-3}\times\eta\times\varepsilon_L,
\eeq
where $\alpha$ is a flavor index, $g_* \simeq 106$ as in the SM, and
$\eta_{\alpha}$ is the efficiency factor of LG under various washout
effects. In the weak washout regime ($\tilde{m} \ll m_*$), we have
$\eta\simeq \tilde{m}^2/m^2_*$, while in the strong washout regime
($\tilde{m}\gg m_*$) we have $\eta\simeq m_*/\tilde{m}$. We use here
the SM value, $C\simeq 12/37$, to characterize the sphaleron effects
that convert L-number into B-number. For the sake of simplicity, we
ignore flavor effects, as they are not changing the order of magnitude
of our results. (For a review of flavor effects see, for example,
\cite{Davidson:2008bu}.)

Similar to standard LG, the asymmetry $\varepsilon_L$ in this case
(with Yukawa coupling $\lambda\epsilon^3$) is given
by~\cite{Covi:1996wh} (with $y_n\equiv M^2_{\beta}/M^2_{\alpha}$)
\beqa \label{eq15}
\varepsilon_{L\alpha\alpha}&\equiv&\frac{\Gamma(N_{\alpha}\rightarrow
LH)-\Gamma(N_{\alpha}\rightarrow \bar{L}H^*)}{\Gamma(N_{\alpha}\rightarrow
LH)+\Gamma(N_{\alpha}\rightarrow \bar{L}H^*)}\nonumber
\\
&=& \displaystyle\sum_{\alpha\neq\beta} \frac{\Im[(\lambda\lambda^{\dagger})^2_{\alpha\beta}]\epsilon^6}{8\pi(\lambda\lambda^{\dagger})_{\alpha\alpha}}\sqrt{y_n}\left[1-(1+y_n)\ln\left(\frac{1+y_n}{y_n}\right)\right]\sim\frac{1}{8\pi}\lambda^2\epsilon^6.
\eeqa
Note that we explicitly kept the $O(1)$ coupling $\lambda$ in
order to demonstrate where the CP violating phase arises. Using the
neutrino mass condition, (\ref{eq12.5}), the RH neutrino decay rate can
be written as
\beq
\Gamma\simeq\frac{\epsilon^6}{8\pi}m_N\sim10^{-13}\frac{m^2_N}{\TeV}.
\eeq
The expansion rate at the time of decay is given by \cite{Kolb:1990vq}
\beq
 H|_{T=m_N}\simeq10^{-15} {m^2_N \over {\TeV}}.
\eeq 
Since $\Gamma\gg H$, the decay is in the strong washout regime. The
baryon asymmetry is therefore
\beq
Y_{\Delta
B}\simeq 10^{-3}\varepsilon_L
\left(\frac{H|_{T=m_N}}{\Gamma}\right)\sim10^{-5}\epsilon^6.
\eeq
Comparing to the observed value, $Y_{\Delta B}\simeq 10^{-10}$, we find
that the following range of parameters lead to successful leptogenesis:
\beq
m_N\sim10^{10}\;\GeV,\qquad 
\epsilon\sim 10^{-1},\qquad 
M\sim10^{16}\;\GeV,\qquad 
\Lambda\sim 10^{15}\;\GeV.
\eeq
These parameters correspond to a high energy LG scenario which gives
the observed values for $m_{\nu}$ and $Y_{\Delta B}$.

\subsection{Dirac-type leptogenesis}

Next we move to study the $T\gg\Lambda$ case. Then the preons are
asymptotically free and we perform all the calculations at the preon
level.  Since we care only about rough estimates we do not include
SU(6)$_C$ radiative corrections. Here we study the L-conserving
model. We get L-number conservation by not including the heavy
Majorana fermion $N$. Below we show that in that case the decay of the
heavy messenger $\Omega$ gives a realization of Dirac-type
LG~\cite{Dick:1999je,Murayama:2002je}.

The idea is as follows. When $T\sim M$, the decay of $\Omega$ and
$\bar{\Omega}$ gives different $L$ and $\bar{L}$ in the final state.
Yet, the decays also generate exactly the same difference between the
number of $A$ and $\bar{A}$. Since $L$ and $A$ carry opposite lepton
numbers, the total lepton number is zero. Yet, each sector ($L$ and
$A$) carry finite and opposite lepton number.  Since the equilibrating
rate is smaller than the expansion rate, the L-number is preserved in
each sector.  When the EW phase transition occurs, sphaleron processes
only affects $L$ and $\bar{L}$, but not $A$ and $\bar{A}$. Thus, the
sphalerons convert part of the L-number stored in the leptons into
B-number. We can end up with positive B-number and negative L-number
in SM sector. Since we only observe the B-number of the universe this
mechanism can be valid.

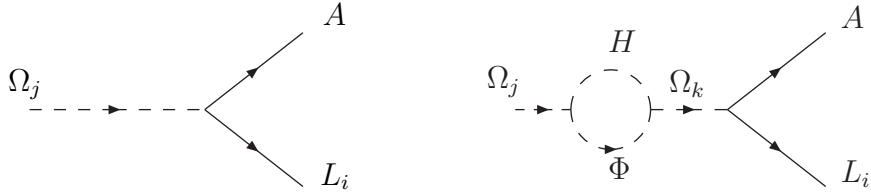
\begin{figure}[t]
\centering
  \begin{picture}(362,119) (-2,1)
    \SetWidth{0.5}
    %
    \DashArrowLine(23,31)(87,31){5}
    \ArrowLine(89,31)(126,60)
    \ArrowLine(89,31)(126,2)
    \Text(15,35)[lb]{\normalsize{{$\Omega_j$}}}
    \Text(134,63)[lb]{\normalsize{{$A$}}}
    \Text(133,1)[lb]{\normalsize{{$L_i$}}}
    \ArrowLine(286,31)(323,60)
    \DashArrowLine(206,31)(227,31){5}
    \DashArrowLine(257,31)(286,31){5}
    \DashArrowArc(242,31)(15,-180,0){5}
    \ArrowLine(286,31)(323,2)
    \DashCArc(242,31)(15,-0,180){5}
    \Text(330,62)[lb]{\normalsize{\Black{$A$}}}
    \Text(196,36)[lb]{\normalsize{\Black{$\Omega_j$}}}
    \Text(330,2)[lb]{\normalsize{\Black{$L_i$}}}
    \Text(265,36)[lb]{\normalsize{\Black{$\Omega_k$}}}
    \Text(242,53)[lb]{\normalsize{\Black{$H$}}}
    \Text(242,5)[lb]{\normalsize{\Black{$\Phi$}}}
  \end{picture}
\caption{The $\Omega$ decay process that gives the L-asymmetry.}
\label{fig:2}
\end{figure}

Specifically we consider the decay $\Omega\rightarrow LA$
(Fig.~\ref{fig:2}). The asymmetry between this decay and its
conjugate process comes  from the interference between the tree level
and the one loop diagrams. It is given by
\beqa \label{Dirac-LG}
\epsilon_{\Omega_j}&\equiv&
\frac{\Gamma(\bar{\Omega}_j\rightarrow\bar{A}\bar{L})-
\Gamma(\Omega_j\rightarrow AL)}
{\Gamma(\bar{\Omega}_j\rightarrow\bar{A}\bar{L})+
\Gamma(\Omega_j\rightarrow AL)} \nonumber \\
&=&
\frac{1}{8\pi}\frac{M^2_j-M^2_{\Phi}}{M^2_{j}-M^2_{k}}
\left(\frac{\tilde{M}_j\tilde{M}_k}{M^2_j}\right)\frac{\Im((Y^{L\dagger }Y^L)_{jk})}{(Y^{L\dagger}Y^L)_{jj}}\sim 
{r^2 \over 8 \pi}, \qquad r\equiv \frac{\tilde{M}}{M}.
\eeqa
Here $j,k=1,2$ and $j\neq k$. $M_j,~M_k,~M_{\Phi}$ are the masses of
$\Omega_i,~\Omega_j,~\Phi$, and we assume $M_{j}\sim M_{k}\sim
M_{\Phi}$ with $M_\Phi < M_j$ such that $\Phi$ can be on-shell in the
loop. Following the convention in Table~\ref{par-count}, we take the
trilinear coupling, $\tilde{M}$, to be real. The CP phase that
contributes to the asymmetry is in $Y^L$.  In half of the parameter
space we end up with negative L-number in the SM sector
and positive L-number in the preon sector.

The natural scale of $\tilde{M}$ is $\tilde{M}\sim M$, that is $r \sim
1$. (Yet, in the following we investigate the allowed parameter space
letting the ratio $r$ to vary.) The main result from
Eq.~(\ref{Dirac-LG}) is that we can get very large lepton
asymmetry. Thus, we have to check if washout effects can reduce the
asymmetry to the observed level.

There are two kinds of washout processes: inverse decays and
scattering that equilibrates the L-number. Here, we would like to demonstrate
that we can get Dirac-LG. Thus, we only try to find some parts of the
parameters space that can produce the observed value of the
asymmetry. We concentrate on the part of the parameter space
where the equilibrating scattering is negligible, that is, where the
equilibrating rate between positive and negative L-numbers is slower
than the expansion of the universe. 

The parameter space where equilibrating scattering is negligible can be
found as follows. First, when $T<M$ the only equilibrating process in
our case is $\bar{A}\bar{L}\rightarrow H\bar{\psi}\bar{\psi}$, coming
from the diagram in Fig.\ref{fig:1}. Its interaction rate can be
estimated as
\beqa\label{equilrate}
R_{eq}|_T\sim|Y^A|^2|Y^L|^2\left(\frac{\tilde{M}}{M}\right)^2\frac{T^7}{M^6}.
\eeqa
Here the $M^{-8}$ factor comes from the masses of virtual $\Omega$
and $\Phi$.  Unlike the original Dirac LG scenario \cite{Dick:1999je}
where $R_{eq}\propto T$, in our case $R_{eq}$ drops much faster than
$H$, that is, $R_{eq} \propto T^2$. Thus, if the equilibrating is
slower than the expansion just when $\Omega$ begins to decay, that is,
\beqa\label{equilcondition}
R_{eq}|_{T=M}\sim|Y^L|^2|Y^A|^2r^2M<H|_{T=M}\sim 10^{-15}\frac{M^2}{\TeV},
\eeqa  
then the equilibrating rate after this is always smaller than the
expansion rate. In that case scattering is very rare and can be
neglected.  That is, by choosing the parameter space satisfying
eq.~(\ref{equilcondition}), we only need to include the inverse decay
for washout effect.

Within this range of parameters we only need to study the effect of
inverse decays. The L-asymmetry is given in eq. (\ref{Dirac-LG}). We
see that for $r>10^{-3}$, the inverse decay must be significant in
order to reduce the asymmetry into the observed value, $Y_{\Delta
B}\sim 10^{-10}$. When including the efficiency factor given by the
strong inverse decay, eq.~(\ref{eq14}), we have the asymmetry
\beq
Y_{\Delta B,\Omega}\simeq 10^{-4}\times r^2\times\left(\frac{H|_{T=M}}{\Gamma_{\Omega}}\right)
\sim 10^{-18}\times r^2\times|Y^L|^{-2}\frac{M}{\TeV}.
\eeq
If the inverse decay lowers the baryon asymmetry to the observed
value, $Y_{\Delta B}\sim 10^{-10}$, the following condition should be
satisfied
\beq\label{eq17}
r^2|Y^L|^{-2}\frac{M}{\TeV}\sim 10^8.
\eeq

We are ready to find a region of the parameter space that gives successful
Dirac-LG. Besides the two constraints
eqs.~(\ref{equilcondition}) and (\ref{eq17})
we also have a constraint from the Dirac neutrino mass
\beqa\label{DiracNuMass}
m_{\nu}=\left(\frac{\tilde{M}}{M}\right)|Y^L||Y^A|\epsilon^3v\sim
10^{-2}\,\eV.
\eeqa
We also require $\epsilon\equiv (\Lambda/M)<10^{-2}$, in order
justify integrating out the heavy scalars. Then, eq.~(\ref{DiracNuMass}) gives
\beqa\label{eq18}
r|Y^L||Y^A|> 10^{-7}.
\eeqa 
Last, we use $|Y^L|, |Y^A| \lsim 1$ in order for perturbation theory
to work. Then, combining eqs.~(\ref{equilcondition}), (\ref{eq17}) and
(\ref{eq18}) we find a representative region in the parameter space
that gives a successful Dirac-type LG:
\beqa\label{eq19}
10^{-3}r&<&|Y^L|<1,\qquad |Y^A|<10^{-4}r^{-2},\qquad |Y^A|<1,\qquad M>10\,\TeV,\nonumber
\\
10^{-7}r^{-4}\TeV&<&M<10^7r^{-2}\,\TeV,\qquad\epsilon<10^{-2}.
\eeqa 
As an example, when $r=1$, the following parameters give a successful
Dirac-LG with strong washout effect
\beqa
|\tilde{M}|=M\qquad M=10\,\TeV\qquad |Y^L|=10^{-3}\qquad |Y^A|=10^{-4}\qquad\epsilon=10^{-2}.
\eeqa
When $r=10^{-3}$, the following parameters give a successful Dirac-LG
with weak washout effect
\beqa
|\tilde{M}|=10^{-3}M\qquad M=10^8\,\TeV\qquad |Y^L|=10^{-3}\qquad |Y^A|=10^{-1}\qquad\epsilon=10^{-2}.
\eeqa

We note that when $r>10^{-2}$, the $\Omega$ mass can be as low as
$10\,\TeV$, which is, much lighter than the Majorana neutrino
mass in the standard LG. The reason that we can get low energy LG
is that the Dirac neutrino mass is not directly related to the lepton
asymmetry. That is, in the composite model the neutrino
mass is suppressed by a factor $(\Lambda/M)^3$. The lepton asymmetry,
however, is proportional to $r$, which is not a very small parameter.
In standard LG, on the contrary, both the neutrino mass and the lepton
asymmetry are proportional to the Yukawa couplings and thus they
cannot be too small.

\section{DISCUSSIONS AND CONCLUSIONS}

We investigated models of composite RH neutrinos. First we find
several UV completions of the models. These full models are not
expected to be unique. They serve as an example that such models can
be constructed. Then we moved on to study leptogenesis in these
models. We find that such models can naturally give leptogenesis. In
particular, we discussed two possibilities corresponding to different
temperatures and particle contents. In the lepton number violating
model we find that they can give standard LG from RH neutrino
decay. In models with lepton number conservation, we find that they can
provide a realization of low energy Dirac LG.  We conclude that the
idea of composite RH neutrino is phenomenologically interesting: it
naturally gives small neutrino masses and successful leptogenesis.

\section*{Acknowledgments}
We thank Josh Berger and Itay Nacshon for helpful discussions. This
research is supported by the NSF grant PHY-0355005.
\\

\appendix
\section{Matching the UV theory to the effective theory}

In this appendix, we obtain the effective Yukawa and L-violating
couplings in eqs. (\ref{eq1}) and (\ref{eq2}) by integrating out the
heavy fields in eqs. (\ref{eq5})-(\ref{eq9}). This gives the relations
between the effective couplings $\lambda$, $h$ and those of the full
theory.

We start from rewriting eqs. (\ref{eq5})-(\ref{eq9}) keeping all the
indices explicitly
\beqa 
&& Y^L_{gi}A_{abm}\sigma^2_{mn}\Omega^{ba\alpha}_{g}L_{i\alpha n}+h.c.,
\\
&& \tilde{M}_{g}\tilde{H}^{\alpha}\Phi^{ab}\Omega_{gba\alpha}+h.c.,
\\
&& Y^A_{ff'}\psi_{fam}\sigma^{2}_{mn}\Phi^{ab}\psi_{f'bn}+h.c.,\label{eq42}
\\
&& Y^N_{k}\epsilon_{abopqr}\epsilon^{opqrst}\epsilon^{uvwxyz}
\Phi^{ab}A_{uvm}\sigma^{2}_{mn}N_{stwxyz,kn}+h.c.,
\\
&& y^N_{ik}H^{\alpha}L_{\alpha i}N_{k}+h.c.
\eeqa
where here the upper indices represent the hermitian conjugate of the
fields. As we can see in eq.~(\ref{eq42}), the antisymmetry in the
spinor and the SU(6)$_C$ indices require $Y^A_{ff'}$ to be
antisymmetric. The indices here are quite cumbersome, and we write
them only when it is necessary in the following calculation.

To obtain the effective Yukawa coupling as an $(\psi A\psi
L\tilde{H})$ vertex, we need to integrate out the heavy $\Omega$ and
$\Phi$ fields in Fig.~\ref{fig:1}a. The $\Omega$ and $\Phi$
related couplings, including their mass terms and three vertices in
the diagram, is
\beq
-M^2\Omega^{\dagger}\Omega-M^2\Phi^{\dagger}\Phi+Y^A\psi\Phi^{\dagger}\psi+
Y^{L\dagger}L^{\dagger}\Omega
A^{\dagger}+\tilde{M}^{\dagger}\Omega^{\dagger}\Phi\tilde{H}+h.c..
\eeq
After integrating $\Omega$ and $\Phi$ out, and using the convention
$|\tilde{M}|=rM$, we obtain
\beq
\frac{1}{M^3}[Y^{L\dagger}rY^A(L^{\dagger}A^{\dagger}\tilde{H})
(\psi^T\psi)+h.c.]
\eeq
Writing the indices explicitly, we can rearrange the fields into a
more transparent form for composite neutrino
\beqa
&&\frac{Y^{L\dagger}_{i}rY^A_{ff'}}{M^3}(L^{*\alpha}_{i
m}\sigma^2_{mn}A^{*ab}_n\tilde{H}_{\alpha})(\psi_{fas}\sigma^2_{st}\psi_{f'bt})+h.c.
\nonumber= \\
&&\frac{Y^{L\dagger}_{i}rY^A_{ff'}}{M^3}(\psi_{fas}\sigma^2_{st}A^{*ab}_m\psi_{f'bt})
\sigma^2_{mn}L^{*\alpha}_{i n}\tilde{H}_{\alpha}+h.c.\nonumber
\equiv\\&&\lambda^{ff',i}\frac{(\psi^T_fA^*\psi_{f'})L^{\dagger}_{i}
\tilde{H}}{M^3}+h.c.,
\eeqa
where
\beq
\label{eq33}\lambda^{ff',i}=Y^{L\dagger}_{i}rY^A_{ff'}\Rightarrow\lambda\sim r|Y^L||Y^A|.
\eeq
Note that the second equality implies that when interchanging $ff'$,
the antisymmetry of $A^{ab}$ and $Y^A_{ff'}$ makes the whole RH
neutrino part invariant. This gives the correct form for $B_{ff'}$,
the massless composite neutrinos.

For the L-violating coupling, eq.~(\ref{eq2}), we need to include the
heavy Majorana fermion $N$. The related couplings in
Fig.~\ref{fig:1}b are:
\beq
-MNN-M^2\Phi^{\dagger}\Phi+Y^{N\dagger}N^{\dagger}A^{\dagger}\Phi A+
Y^{A}\psi\Phi^{\dagger}\psi+h.c..
\eeq
After integrating out $N$ and $\Phi$, we obtain
\beq
\frac{(Y^AY^{N\dagger})^2}{4M^5}(\psi^T\psi
A^*)(A^{\dagger}\psi^T\psi)+h.c..
\eeq
Writing this in a form that is best for studying composite neutrinos, we have
\beqa
&&\frac{(Y^A_{ff'}Y^{N\dagger})(Y^A_{gg'}Y^{N\dagger})}{4M^5}
(\psi_{fm}\sigma^2_{mn}\psi_{f'n}
A^{\dagger}_o)\sigma^2_{op}(A_p^{*}\psi_{gs}\sigma^2_{st}\psi_{g't}) =
\nonumber \\ &&
\frac{(Y^A_{ff'}Y^{N\dagger})(Y^A_{gg'}Y^{N\dagger})}{4M^5}(\psi_{fm}\sigma^2_{mn}A^{\dagger}_o\psi_{f'n}
)^T\sigma^2_{op}(\psi_{gs}\sigma^2_{st}A_p^{*}\psi_{g't})\equiv 
\nonumber \\ &&
h^{ff',gg'}\frac{(\psi^T_{f} A^{\dagger}\psi_{f'})
(\psi^T_{g} A^{*}\psi_{g'})}{M^5},
\eeqa
where
\beq
\label{eq34}h^{ff',gg'}=\frac{1}{4}(Y^A_{ff'}Y^{N\dagger})(Y^A_{gg'}Y^{N\dagger})
\quad\Rightarrow\quad
h\sim|Y^N|^2|Y^A|^2.
\eeq

\section{Calculation of $\mu \to e \gamma$}

\begin{figure}[t]
\centering
\begin{picture}(386,66) (24,-59)
    \SetWidth{0.5}
    \ArrowLine(31,-53)(69,-38)
    \ArrowLine(69,-38)(107,-53)
    \Photon(69,-8)(69,-38){3.5}{3}
    \Text(79,-9)[lb]{\normalsize{\Black{$\gamma$}}}
    \Text(24,-50)[lb]{\normalsize{\Black{$L$}}}
    \Text(109,-50)[lb]{\normalsize{\Black{$L$}}}
    \Text(60,-70)[lb]{\normalsize{\Black{$(a)$}}}
    \DashArrowLine(166,-54)(204,-39){5}
    \DashArrowLine(204,-39)(242,-54){5}
    \Photon(204,-9)(204,-39){3.5}{3}
    \Text(214,-11)[lb]{\normalsize{\Black{$\gamma$}}}
    \Text(160,-50)[lb]{\normalsize{\Black{$\Omega$}}}
    \Text(236,-50)[lb]{\normalsize{\Black{$\Omega$}}}
    \Text(181,-60)[lb]{\footnotesize{\Black{$p$}}}
    \Text(218,-60)[lb]{\footnotesize{\Black{$p'$}}}
    \Text(197,-70)[lb]{\normalsize{\Black{$(b)$}}}
    \ArrowLine(299,-55)(337,-40)
    \DashArrowLine(337,-39)(337,-8){5}
    \ArrowLine(374,-55)(337,-40)
    \Text(347,-12)[lb]{\normalsize{\Black{$\Omega$}}}
    \Text(291,-50)[lb]{\normalsize{\Black{$L$}}}
    \Text(380,-50)[lb]{\normalsize{\Black{$A$}}}
    \Text(332,-70)[lb]{\normalsize{\Black{$(c)$}}}
  \end{picture}
\centering
\begin{picture}(422,100) (65,-52)
    \SetWidth{0.5}
    \ArrowLine(76,-44)(121,-44)
    \ArrowLine(196,-44)(241,-44)
    \ArrowLine(286,-44)(331,-44)
    \ArrowLine(406,-44)(451,-44)
    \DashCArc(158.5,-52.43)(38.44,12.67,167.33){5}
    \DashCArc(368.5,-52.43)(38.44,12.67,167.33){5}
    \Photon(384,-17)(422,-2){3.5}{4}
    \Photon(104,-44)(134,1){3.5}{5}
    \ArrowLine(196,-44)(121,-44)
    \ArrowLine(406,-44)(331,-44)
    \Text(65,-40)[lb]{\normalsize{\Black{$\mu$}}}
    \Text(275,-40)[lb]{\normalsize{\Black{$\mu$}}}
    \Text(134,8)[lb]{\normalsize{\Black{$\gamma$}}}
    \Text(427,6)[lb]{\normalsize{\Black{$\gamma$}}}
    \Text(155,-8)[lb]{\normalsize{\Black{$\Omega$}}}
    \Text(363,-7)[lb]{\normalsize{\Black{$\Omega$}}}
    \Text(109,-19)[lb]{\footnotesize{\Black{$q$}}}
    \Text(400,-1)[lb]{\footnotesize{\Black{$q$}}}
    \Text(155,-25)[lb]{\footnotesize{\Black{$k$}}}
    \Text(366,-25)[lb]{\footnotesize{\Black{$k$}}}
    \Text(403,-33)[lb]{\footnotesize{\Black{$(k-q)$}}}
    \Text(355,-42)[lb]{\footnotesize{\Black{$(p-k)$}}}
    \Text(143,-42)[lb]{\footnotesize{\Black{$(p'-k)$}}}
    \Text(246,-41)[lb]{\normalsize{\Black{$e$}}}
    \Text(457,-41)[lb]{\normalsize{\Black{$e$}}}
    \Text(157,-57)[lb]{\normalsize{\Black{$A$}}}
    \Text(367,-57)[lb]{\normalsize{\Black{$A$}}}
    \Text(82,-55)[lb]{\footnotesize{\Black{$p$}}}
    \Text(222,-55)[lb]{\footnotesize{\Black{$p'$}}}
    \Text(306,-55)[lb]{\footnotesize{\Black{$p$}}}
    \Text(426,-55)[lb]{\footnotesize{\Black{$p'$}}}
  \end{picture}
\caption{In the upper part are the vertices we use in the calculation:
($a$) $-ie\gamma^{\mu}$ ($b$) $-ie(p+p')^{\mu}$ ($c$) $iY^L$.
The lower part are the kinematics we use in the calculation. The case with
the photon going out from $e$ is not shown, since we can obtain
the result directly from the first diagram.}
\label{fig:5}
\end{figure}
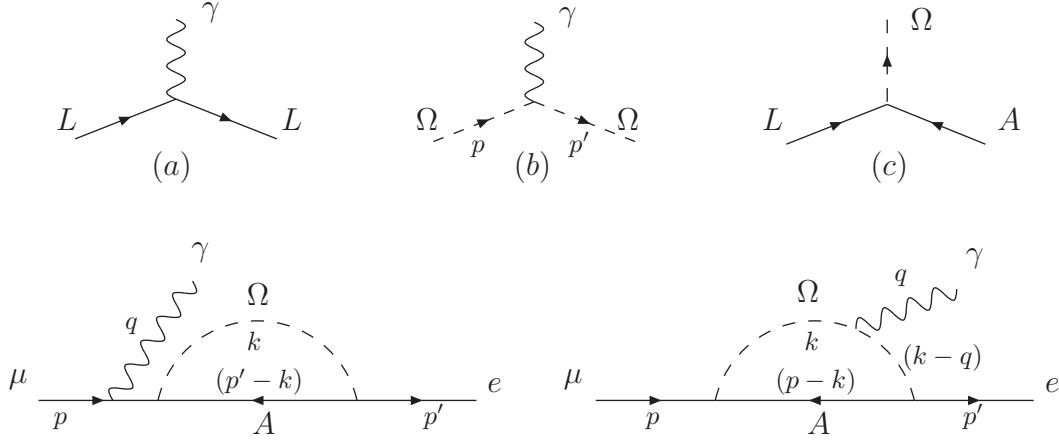

In this appendix, we calculate the bounds on $M$ given by the lepton
flavor violating (LFV) process $\mu\rightarrow e\gamma$.  The vertices
and the kinematics of the LFV process are shown in
Fig.~\ref{fig:5}. 

Throughout the calculation, we neglect the mass of the out-going
electron.  We first evaluate the amplitude of the diagram where the
photon coming from the external muon.  This diagram scales as the
electron mass and thus vanish in the limit of massless electron.

Explicitly the diagram gives
\beqa
M_{\mu\rightarrow\gamma}&=&\bar{u}_{e_R}(-iY^*_L)\int\frac{d^4k}{(2\pi)^4}\frac{i}{k^2-M^2}\frac{i(\displaystyle{\not}p'-\displaystyle{\not}k)}{(p'-k)^2}(iY_L)\frac{i(\displaystyle{\not}p'+m_{\mu})}{(p'^2-m^2_{\mu})}(-ie\displaystyle{\not}\varepsilon)u_{\mu}\nonumber
\\
&=&-e|Y^L|^2\bar{u}_{e_R}\left[\int\frac{d^4k}{(2\pi)^4}\frac{\displaystyle{\not}p'-\displaystyle{\not}k}{(k^2-M^2)(p'-k)^2}\right]\frac{\displaystyle{\not}p'+m_{\mu}}{p'^2-m^2_{\mu}}\displaystyle{\not}\varepsilon
u_{\mu}.
\eeqa
Here $M$, $m_{\mu}$, $m_e$ are the masses of $\Omega$, $\mu$, $e$, we
use $p'\equiv(p-q)$, and $\varepsilon^{\mu}$ is the polarization of
the outgoing photon. Integrating out the loop momentum and doing the
dimensional regularization, we get the amplitude as
\beqa
M_{\mu\rightarrow\gamma}
=\frac{-ie|Y^L|^2}{32\pi^2}\bar{u}_{e_R}\left(\displaystyle{\not}p'\frac{\displaystyle{\not}p'+m_{\mu}}{p'^2-m^2_{\mu}}\displaystyle{\not}\epsilon\right)\left(\frac{2}{\epsilon}-\gamma+\ln(4\pi)+\frac{1}{2}-\ln M^2\right)u_{\mu}.
\eeqa
Here $\gamma$ is the Euler-Mascheroni constant, $\epsilon\equiv 4-d$
and we take $d\rightarrow 4$ for the finite terms. We use the
condition of transverse polarization 
\beqa\label{polarization}
\varepsilon_{\mu}q^{\mu}=0,\qquad \varepsilon_{\mu}p^{\mu}=0,\qquad\varepsilon_{\mu}p'^{\mu}=0.
\eeqa
Then, we see that the diagram vanishes, that is,
$M_{\mu\rightarrow\gamma}=0$.

The amplitude of the diagram where the external photon is
emitted by the electron can be written as
\beqa
M_{e\rightarrow\gamma}=\bar{u}_{e_R}(-ie\displaystyle{\not}\varepsilon)\frac{i(\displaystyle{\not}p)}{(p^2)}(-iY^*_L)\int\frac{d^4k}{(2\pi)^4}\frac{i}{k^2-M^2}\frac{i(\displaystyle{\not}p-\displaystyle{\not}k)}{(p-k)^2}(iY_L)u_{\mu}\nonumber
\\
=-e|Y^L|^2\bar{u}_e\displaystyle{\not}\varepsilon\frac{\displaystyle{\not}p}{p^2}\left[\int\frac{d^4k}{(2\pi)^4}\frac{\displaystyle{\not}p-\displaystyle{\not}k}{(k^2-M^2)(p-k)^2}\right]
u_{\mu}. 
\eeqa
Integrating out the loop momentum and doing the regularization, this gives
\beqa\label{eq39}
M_{e\rightarrow\gamma}
&=&\frac{-ie|Y^L|^2}{32\pi^2}\bar{u}_{e_R}{\not}\varepsilon\left(\frac{{\not}p}{m_{\mu}^2}{\not}p\right)\left[\frac{2}{\epsilon}-\gamma+\ln(4\pi)+\frac{1}{2}-\ln M^2+\frac{1}{3}\left(\frac{m_{\mu}}{M}\right)^2\right]
u_{\mu}\nonumber
\\
&=&\frac{-ie|Y^L|^2}{32\pi^2}\left[\frac{2}{\epsilon}-\gamma+\ln(4\pi)+\frac{1}{2}-\ln M^2+\frac{1}{3}\left(\frac{m_{\mu}}{M}\right)^2\right]
\varepsilon_{\nu}\bar{u}_{e_R}\gamma^{\nu} u_{\mu_{R}}.\label{ElectronResult}
\eeqa
when keeping terms up to order $O(m_{\mu}^2/M^2)$.

For the case with the photon coming out from the internal $\Omega$
(see Fig.\ref{fig:5}), the amplitude is
\beqa
M_{\Omega\rightarrow\gamma}&=&\bar{u}_{e_R}(-iY^*_L)\varepsilon_{\nu}\int\frac{d^4k}{(2\pi)^4}\frac{i}{(k-q)^2-M^2}(-ie(2k-q)^{\nu})\frac{i}{(k^2-M^2)}\frac{i(\displaystyle{\not}p-\displaystyle{\not}k)}{(p-k)^2}(iY_L)u_{\mu}\nonumber
\\
&=&-e|Y^L|^2\varepsilon_{\nu}\bar{u}_{e_R}\left[\int\frac{d^4k}{(2\pi)^4}\frac{(2k-q)^{\nu}({\not}p-{\not}k)}{((k-q)^2-M^2)(k^2-M^2)(p-k)^2}\right]u_{\mu}.
\eeqa
Integrating out the loop momentum, taking $m_e=0$ and using the transverse polarization condition, eq.~(\ref{polarization}), we get the amplitude when keeping the terms up to $O(\frac{m_{\mu}^2}{M^2})$
\beqa\label{eq41}
M_{\Omega\rightarrow\gamma}
=\frac{ie|Y^L|^2}{32\pi^2}\left[\frac{2}{\epsilon}-\gamma+\ln(4\pi)+\frac{1}{2}-\ln M^2+\frac{1}{6}\left(\frac{m_{\mu}}{M}\right)^2\right]\varepsilon_{\nu}\bar{u}_{e_R}\gamma^{\nu}u_{\mu_R}
.\eeqa
Combining the three diagrams, we have
\beq\label{eq41.5}
M_{\mu\rightarrow e\gamma}=-\frac{ie|Y^L|^2}{192\pi^2}\left(\frac{m_{\mu}}{M}\right)^2\bar{u}_{e_R}\displaystyle{\not}\varepsilon u_{\mu_R}.
\eeq
Using $m_e=0$ and eq.~(\ref{polarization}), we can write the result into the well known dipole operator
\beq
\frac{ie|Y^L|^2}{768\pi^2}\left(\frac{m_{\mu}}{M^2}\right)\bar{e}_R\sigma_{\mu\nu}F^{\mu\nu}\mu_L
.\eeq
Averaging the incoming muon spin, the amplitude square becomes
\beq
<|M|^2>_{spin}=-\frac{e^2|Y^L|^4}{
2\times 192^2\pi^4}(\frac{m_{\mu}}{M})^4Tr[{\not}p_e\gamma^{\mu}({\not}p_{\mu})\gamma_{\mu}]
=\frac{\alpha|Y^L|^4}{96^2\pi^3}\left(\frac{m^6_{\mu}}{M^4}\right).
\eeq
This gives the decay rate
\beqa
\Gamma(\mu\rightarrow\gamma e)=\frac{1}{32\pi^2}<|M|^2>_{spin}\frac{|q|}{m^2_{\mu}}\int
d\Omega=\frac{\alpha|Y^L|^4}{
768^{2}\pi^4}\frac{m^5_{\mu}}{M^4}.
\eeqa
Comparing to the total muon decay rate $\frac{G^2_{F}m^5_{\mu}}{192\pi^3}$
, this gives the branching ratio
\beqa\label{muegamma}
Br(\mu\rightarrow\gamma e)=\frac{\alpha |Y^L|^4}{3072\pi G^2_{F}M^4}.
\eeqa
Comparing to the LFV bound
today $Br(\mu\rightarrow eX)<10^{-11}$ \cite{Amsler:2008zz}, we have
\beq
M> 10|Y^L|\,\TeV.
\eeq

\section{Coherent muon-electron conversion}

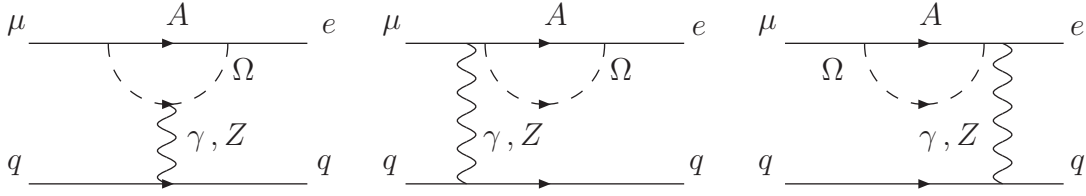
\begin{figure}[t]
\centering
 \begin{picture}(427,78) (39,-58)
    \SetWidth{0.5}
    \ArrowLine(47,-4)(152,-4)
    \DashArrowArc(99.5,-4.5)(22.51,178.73,361.27){5}
    \Photon(99,-27)(99,-57){3.5}{3}
    \ArrowLine(47,-57)(152,-57)
    \ArrowLine(189,-4)(294,-4)
    \ArrowLine(332,-4)(437,-4)
    \DashArrowArc(241.5,-4.5)(22.51,178.73,361.27){5}
    \DashArrowArc(384.5,-4.5)(22.51,178.73,361.27){5}
    \ArrowLine(189,-57)(294,-57)
    \ArrowLine(332,-57)(437,-57)
    \Photon(414,-4)(414,-57){3.5}{5}
    \Photon(212,-4)(212,-57){3.5}{5}
    \Text(39,-1)[lb]{\normalsize{\Black{$\mu$}}}
    \Text(181,-1)[lb]{\normalsize{\Black{$\mu$}}}
    \Text(323,-1)[lb]{\normalsize{\Black{$\mu$}}}
    \Text(158,0)[lb]{\normalsize{\Black{$e$}}}
    \Text(298,-1)[lb]{\normalsize{\Black{$e$}}}
    \Text(441,-1)[lb]{\normalsize{\Black{$e$}}}
    \Text(39,-54)[lb]{\normalsize{\Black{$q$}}}
    \Text(156,-54)[lb]{\normalsize{\Black{$q$}}}
    \Text(297,-54)[lb]{\normalsize{\Black{$q$}}}
    \Text(181,-54)[lb]{\normalsize{\Black{$q$}}}
    \Text(323,-54)[lb]{\normalsize{\Black{$q$}}}
    \Text(441,-54)[lb]{\normalsize{\Black{$q$}}}
    \Text(124,-18)[lb]{\normalsize{\Black{$\Omega$}}}
    \Text(267,-18)[lb]{\normalsize{\Black{$\Omega$}}}
    \Text(347,-18)[lb]{\normalsize{\Black{$\Omega$}}}
    \Text(107,-45)[lb]{\normalsize{\Black{$\gamma\,,Z$}}}
    \Text(219,-45)[lb]{\normalsize{\Black{$\gamma\,,Z$}}}
    \Text(384,-45)[lb]{\normalsize{\Black{$\gamma\,,Z$}}}
    \Text(99,3)[lb]{\normalsize{\Black{$A$}}}
    \Text(242,3)[lb]{\normalsize{\Black{$A$}}}
    \Text(384,3)[lb]{\normalsize{\Black{$A$}}}
  \end{picture}
\caption{$\mu-e$ conversion in nuclei emitted by photon and $Z$.}
\label{fig:6}
\end{figure}

In this appendix we estimate the bounds from the LFV process of
coherent muon-electron conversion (Fig.~\ref{fig:6}).  For a review of
the coherent conversion and how it can be used to put bounds on new
physics, see~\cite{Czarnecki:1998iz, Kitano:2002mt} for example.

Our goal is to find the bound on $M$ by comparing the
theoretical expression with experimental data. Here we use the
general result derived in \cite{Czarnecki:1998iz} for the theoretical
branching ratio.  The low energy effective Hamiltonian
is~\cite{Czarnecki:1998iz}
\beqa\label{conversion1}
H&=&-\bar{e}\tilde{O}\mu+h.c.\nonumber
\\
\tilde{O}&=&-\sqrt{4\pi\alpha}\left[
\gamma_{\alpha}(f_{E0}-f_{M0}\gamma_5)\frac{q^2}{m_{\mu}^2}+
i\sigma_{\alpha\beta}\frac{q^{\beta}}{m_{\mu}}(f_{M_1}+f_{E1}\gamma_5)\right]A^{\alpha}(q)
+\frac{G_F}{\sqrt{2}}\gamma_{\alpha}(a-b\gamma_5)J^{\alpha}\nonumber
\\
J^{\alpha}&=&\bar{u}\gamma^{\alpha}u+c_d\bar{d}\gamma^{\alpha}d
\eeqa
and the final result of the conversion rate is 
\beqa
\label{eq43} w_{conv}&=&3\times 10^{23}(w^{(1)}_{conv}+w^{(2)}_{conv})\sec^{-1},\nonumber
\\
\label{eq44}w^{(1)}_{conv}&=&\left|f_{E0}I_p-\frac{G_F}{\sqrt{2}}\frac{m_{\mu}^2}{4\pi
Z\alpha}a(Z(2+c_d)I_p+N(1+2c_d)I_n)+f_{M1}I_{34}\right|^2,\nonumber
\\
\label{eq45}w^{(2)}_{conv}&=&\left|f_{M0}I_p-\frac{G_F}{\sqrt{2}}\frac{m_{\mu}^2}{4\pi
Z\alpha}b(Z(2+c_d)I_p+N(1+2c_d)I_n)+f_{E1}I_{34}\right|^2,
\eeqa
where
\beq
I_p=-(I^p_1+I^p_2),\qquad
I_n=-(I^n_1+I^n_2),\qquad
I_{34}=I_3+I_4.
\eeq
Here $q$ represents the photon momentum, and the terms
containing $A^{\alpha}$ in the Hamiltonian describe the transition that is mediated by a photon. The $I$'s in the last part are coefficients for various elements including the proton-neutron distribution function and the EM field inside the nucleus. They have been calculated in \cite{Kitano:2002mt} for various materials.

We are ready to use these results in the composite model. The rate of
$\mu N\rightarrow eN$ arising from the preon sector is given by the
six diagrams in Fig.~\ref{fig:6}. Doing the same calculation as in
appendix B but allowing the out-going photon to be off-shell, the
coefficients in eq.~(\ref{conversion1}) are of order
\beqa
f_{E0}\sim -f_{M0}\sim f_{M_1}\sim -f_{E_1}\sim a\sim b\sim
\frac{|Y^L|^2}{768 \pi^2}\frac{m^2_{\mu}}{M^2},\qquad c_d\sim 1.
\eeqa

Given these coefficients and the $I$'s calculated in \cite{Czarnecki:1998iz,Kitano:2002mt} (which are of order $10^{-1}\,\GeV^{-\frac{1}{2}}$), the conversion rate with target $^{48}_{22}Ti$ can be estimated as:
\beqa
w_{conv}\sim 10^{14}|Y^L|^4\left(\frac{m_{\mu}}{M}\right)^4\sec^{-1}.
\eeqa
Comparing to the experimental total muon capture rate $w(Ti)_{cap}=2.6\times 10^{6}\,\sec^{-1}$ \cite{Suzuki:1987jf}, this gives the branching ratio of the conversion as
\beq\label{eq47}
Br(\mu\rightarrow e,
Ti)\equiv\frac{w_{conv}}{w_{cap}}=10^{8}|Y^L|^4\left(\frac{m_{\mu}}{M}\right)^4.
\eeq
Comparing to the experimental limit $Br(\mu\rightarrow e)<1.7\times 10^{-12}$ \cite{Kaulard:1998rb}, this gives
\beqa
M>10|Y^L|\,\TeV.
\eeqa



\begin{thebibliography}{99}

\bibitem{Fukugita:1986hr}
  M.~Fukugita and T.~Yanagida,
  Phys.\ Lett.\  B {\bf 174}, 45 (1986).

\bibitem{Davidson:2008bu}
  S.~Davidson, E.~Nardi and Y.~Nir,
  arXiv:0802.2962 [hep-ph].

\bibitem{ArkaniHamed:1998pf}
  N.~Arkani-Hamed and Y.~Grossman,
  Phys.\ Lett.\  B {\bf 459}, 179 (1999)
  [arXiv:hep-ph/9806223].

\bibitem{Okui:2004xn}
  For searching composite neutrinos in CMB, T.~Okui,
  JHEP {\bf 0509}, 017 (2005)
  [arXiv:hep-ph/0405083].


\bibitem{Dimopoulos:1980hn}
  S.~Dimopoulos, S.~Raby and L.~Susskind,
  Nucl.\ Phys.\  B {\bf 173}, 208 (1980).

\bibitem{Amsler:2008zz}
  C.~Amsler {\it et al.}  [Particle Data Group],
  Phys.\ Lett.\  B {\bf 667}, 1 (2008).

\bibitem{Kaulard:1998rb}
  J.~Kaulard {\it et al.}  [SINDRUM II Collaboration],
  Phys.\ Lett.\  B {\bf 422}, 334 (1998).

\bibitem{Cyburt:2004yc}
  An incomplete list. R.~H.~Cyburt, B.~D.~Fields, K.~A.~Olive and E.~Skillman,
  Astropart.\ Phys.\  {\bf 23}, 313 (2005)
  [arXiv:astro-ph/0408033]; U.~Seljak, A.~Slosar and P.~McDonald,
  JCAP {\bf 0610}, 014 (2006)
  [arXiv:astro-ph/0604335]; J.~Dunkley {\it et al.}  [WMAP Collaboration],
  arXiv:0803.0586 [astro-ph].

\bibitem{Kolb:1990vq}
  E.~W.~Kolb and M.~S.~Turner,
  Front.\ Phys.\  {\bf 69}, 1 (1990); 
  S.~Dodelson,
{\it  Amsterdam, Netherlands: Academic Pr. (2003) 440 p}


\bibitem{Covi:1996wh}
  L.~Covi, E.~Roulet and F.~Vissani,
  Phys.\ Lett.\  B {\bf 384}, 169 (1996)
  [arXiv:hep-ph/9605319].

\bibitem{Dick:1999je}
  K.~Dick, M.~Lindner, M.~Ratz and D.~Wright,
  Phys.\ Rev.\ Lett.\  {\bf 84}, 4039 (2000)
  [arXiv:hep-ph/9907562].

\bibitem{Murayama:2002je}
  H.~Murayama and A.~Pierce,
  Phys.\ Rev.\ Lett.\  {\bf 89}, 271601 (2002)
  [arXiv:hep-ph/0206177].

\bibitem{Czarnecki:1998iz}
  A.~Czarnecki, W.~J.~Marciano and K.~Melnikov,
  AIP Conf.\ Proc.\  {\bf 435}, 409 (1998)
  [arXiv:hep-ph/9801218].

\bibitem{Kitano:2002mt}
  R.~Kitano, M.~Koike and Y.~Okada,
  Phys.\ Rev.\  D {\bf 66}, 096002 (2002)
  [Erratum-ibid.\  D {\bf 76}, 059902 (2007)]
  [arXiv:hep-ph/0203110].

\bibitem{Suzuki:1987jf}
  T.~Suzuki, D.~F.~Measday and J.~P.~Roalsvig,
  Phys.\ Rev.\  C {\bf 35}, 2212 (1987).




\end{thebibliography}
\end{document}